\documentclass[aps,prc,twocolumn,superscriptaddress,nofootinbib,longbibliography,floatfix,10pt]{revtex4-1}

\usepackage{graphicx}
\usepackage{dcolumn}
\usepackage{bm}
\usepackage{morefloats}
\usepackage{multirow}
\usepackage{amssymb}
\usepackage{amsmath}
\usepackage{xcolor}
\usepackage{longtable}
\usepackage{fix-cm}
\usepackage{mathptmx} 
\usepackage[T1]{fontenc}
\usepackage[colorlinks,allcolors=blue]{hyperref}
\makeatletter
\def\NAT@def@citea{\def\@citea{\NAT@separator}}
\makeatother

\begin{document}

\title{Quantal diffusion description of isotope production by multinucleon transfer mechanism in ${}^{48} \text{Ca}+{}^{238} \text{U}$ collisions}

\author{S. Ayik}\email{ayik@tntech.edu}
\affiliation{Physics Department, Tennessee Technological University, Cookeville, TN 38505, USA}
\author{M. Arik}
\affiliation{Physics Department, Middle East Technical University, 06800 Ankara, Turkey}
\author{E. C. Karanfil}
\affiliation{Physics Department, Middle East Technical University, 06800 Ankara, Turkey}
\author{O. Yilmaz}
\affiliation{Physics Department, Middle East Technical University, 06800 Ankara, Turkey}
\author{B. Yilmaz}
\affiliation{Physics Department, Faculty of Sciences, Ankara University, 06100 Ankara, Turkey}
\author{A. S. Umar}
\affiliation{Department of Physics and Astronomy, Vanderbilt University, Nashville, TN 37235, USA}

\date{\today}

\begin{abstract}
As an extension of previous work, we calculate the production cross-section of heavy neutron-rich isotopes by employing the quantal diffusion description to ${}^{48} \text{Ca} + {}^{238} \text{U}$ collisions.  The quantal diffusion is deduced from stochastic mean-field approach, and transport properties are determined in terms of time-dependent single-particle wave functions of the time-dependent Hartree-Fock (TDHF) theory. As a result, the approach allows for prediction of production cross-sections without any adjustable parameters.  The secondary cross-sections by particle emission are calculated with the help of the statistical GEMINI++ code.
\end{abstract}

\maketitle

\section{Introduction}
\label{intro}
Extensive experimental investigations of the multi-nucleon transfer process have been done in heavy-ion collisions with actinide targets at near barrier energies~\cite{kozulin2012,kozulin2014b,kratz2013,watanabe2015,devaraja2015,desai2019}, and more investigation are currently in progress.  These studies may provide an efficient mechanism for production of heavy neutron-rich isotopes, which may not otherwise be possible in fusion, fission, and fragmentation reactions. In quasi-fission reactions, colliding nuclei stick together in a di-nuclear configuration. During long contact times, large number of  nucleons are transfered between projectile-like and target-like nuclei.  

In theoretical studies of multi-nucleon transfer mechanism a number of macroscopic models have been employed including the di-nuclear system model~\cite{adamian2003,adamian2010,adamian2010c} and Langevin-type stochastic models~\cite{zagrebaev2008c,zagrebaev2011,zagrebaev2012,karpov2017,saiko2019}. These phenomenological models with a number of adjustable parameters provide qualitative and partly semi-quantitative description of the reaction mechanism. 

In order to provide a more reliable predictive capability, it is highly desirable to develop a microscopic description without any adjustable parameters. The time-dependent Hartree-Fock (TDHF) theory with effective interactions (effective energy density functional) provides a microscopic description for nuclear dynamics at low energies where the Pauli blocking is very effective~\cite{simenel2012,nakatsukasa2016,oberacker2014,umar2015a,simenel2020}.
The applicability of TDHF to study quasi-fission for a wide selection of systems is well established~\cite{wakhle2014,oberacker2014,umar2015a,hammerton2015,umar2016,sekizawa2016,guo2018d,zheng2018,godbey2019,simenel2020} (see~\cite{simenel2012,simenel2018,sekizawa2019,stevenson2019} for recent reviews of TDHF applications to  heavy-ion reactions).
 However, the TDHF theory has a severe limitation: it can only describe the most probable dynamical path of the collision dynamics with small fluctuations around it. It describes the mean kinetic energy loss due to one body dissipation rather well but it cannot describe the large dispersions of mass and charge distribution of the fragments. Particle projection method applied to TDHF clearly demonstrates that dispersion of a few nucleon transfers are described reasonable well but it falls short for a large number of transfers~\cite{simenel2010,sekizawa2016}. 

This suggests that an improvement of the TDHF theory beyond the mean-afield approximation is required.  The time-dependent random phase approximation of Balian and V\'en\'eroni provided a significant improvement beyond the mean field approximation for the dispersions of the one-body observables~\cite{balian1985,williams2018,godbey2020}. The stochastic mean-field (SMF) approach provides further improvement of the TDHF beyond the mean field approximation~\cite{ayik2008,lacroix2014}. In Sec.~\ref{descr} we briefly describe the quantal diffusion description of multi-nucleon transfer based on the SMF approach. Section~\ref{results} presents the result of calculations of isotope cross-section produced in ${}^{48} \text{Ca}+{}^{238} \text{U}$ collisions at $E_{\text{c.m.}} =193$~MeV. In Sec.~\ref{conc}, conclusions are given.

\section{Quantum Diffusion Description of Multinucleon Transfer}
\label{descr}
In TDHF theory a unique single-particle density matrix is calculated with a given initial condition. On the other hand, in the SMF approach to the mean-field theory an ensemble of single-particle density matrices are generated by incorporating the fluctuations of the initial state. The single particle density matrix in each event is determined by the TDHF equations with the self-consistent Hamiltonian of that event. In each event of the SMF approach, fluctuations of the random element of the initial density matrices are determined by Gaussian distributions with variances specified by the requirement that ensemble average of dispersions of one-body observables matches the quantal expressions in the mean-field approach. 

When a di-nuclear structure is maintained in the collision dynamics, we do not need to generate an ensemble of mean-field events.  In this case, it is possible to develop much easier transport description in terms of Langevin transport equations for relevant macroscopic variables by the geometric projection procedure of the SMF approach with the help of the window dynamics. For details of the quantal diffusion description and the window dynamics we refer to Refs.~\cite{ayik2017,ayik2018,yilmaz2018,ayik2019,sekizawa2020,ayik2020}.  For describing nucleon diffusion mechanism, we consider neutron number and proton number of the projectile-like fragments as relevant macroscopic variables. We can determine the neutron $N_{1}^{\lambda}(t)$ and proton $Z_{1}^{\lambda}(t)$ numbers of the projectile-fragments in an event $\lambda$ by integrating the particle density on the left side or the right side of the window, according to the window dynamics, 
\begin{align}\label{eq1}
\left(\begin{array}{c} N_{1}^{\lambda}(t) \\ Z_{1}^{\lambda} (t) \end{array}\right)=\int d^{3} r\,&\Theta [(x-x_{0} )\cos \theta +(y-y_{0} )\sin \theta]\nonumber\\
&\times\left(\begin{array}{c} {\rho _{n}^{\lambda } (\vec{r},t)} \\ {\rho _{p}^{\lambda } (\vec{r},t)} \end{array}\right)\;,
\end{align} 
where the quantity
\begin{align}\label{eq2}
\rho _{\alpha }^{\lambda } (\vec{r},t)=\sum _{ij\in \alpha }\Phi _{j}^{*\alpha }  (\vec{r},t;\lambda )\rho _{ji}^{\lambda } \Phi _{i}^{\alpha } (\vec{r},t;\lambda )\;,
\end{align} 
denotes the neutron and proton number $(\alpha=n, p)$ densities in the event $\lambda$ of the ensemble of the single-particle density matrices. In this expression, $x_{0} (t)$ and $y_{0} (t)$ are the coordinates of the window center relative to the origin of the center of mass frame, $\theta (t)$ is the smaller angle between the orientation of the symmetry axis of the di-nuclear system and the beam direction. Neutron and proton numbers of the projectile-like fragments (or target-like fragments) fluctuate from one event to another, and these numbers can be decomposed as $N_{1}^{\lambda } (t)=N_{1} (t)+\delta N_{1}^{\lambda } (t)$ and $Z_{1}^{\lambda } (t)=Z_{1} (t)+\delta Z_{1}^{\lambda } (t)$. Here, $N_{1} (t)$ and $Z_{1} (t)$ are the mean values determined by the mean-field description of the TDHF theory. According to the quantal diffusion approach, fluctuations of the neutron $\delta N_{1}^{\lambda } (t)$ and the proton $\delta Z_{1}^{\lambda } (t)$ numbers evolve according to the coupled Langevin equations,
\begin{align}\label{eq3}
\frac{d}{dt} \left(\begin{array}{c} {\delta Z_{1} (t)} \\ {\delta N_{1} (t)} \end{array}\right)=&\left(\begin{array}{c} {\frac{\partial v_{p} }{\partial Z_{1} } \left(Z_{1}^{\lambda } -Z_{1} \right)+\frac{\partial v_{p} }{\partial N_{1} } \left(N_{1}^{\lambda } -N_{1} \right)} \\ {\frac{\partial v_{n} }{\partial Z_{1} } \left(Z_{1}^{\lambda } -Z_{1} \right)+\frac{\partial v_{n} }{\partial N_{1} } \left(N_{1}^{\lambda } -N_{1} \right)} \end{array}\right)\nonumber\\
&+\left(\begin{array}{c} {\delta v_{p}^{\lambda } (t)} \\ {\delta v_{n}^{\lambda } (t)} \end{array}\right)\;,
\end{align} 
where quantities $v_{\alpha }^{\lambda } (t)=v_{\alpha } (t)+\delta v_{\alpha }^{\lambda } (t)$ are the drift coefficients of neutrons and protons with the mean values and the fluctuating parts are expressed by $v_{\alpha} (t)$ and $\delta v_{\alpha }^{\lambda } (t)$, with $\alpha$ denoting neutron and proton labels. The linear limit of Langevin description presented here provides a good approximation when the driving potential energy is nearly harmonic around the mean values of the mass and charge asymmetry. The mean values of drift coefficients are extracted from TDHF, and their derivatives are evaluated at the mean values. The explicit expressions of the stochastic parts of drift coefficients $\delta v_{\alpha }^{\lambda } (t)$ can be found in Ref.~\cite{ayik2018}.

\subsection{Quantal Diffusion Coefficients}
\begin{figure}[!t]
\includegraphics*[width=8.6cm]{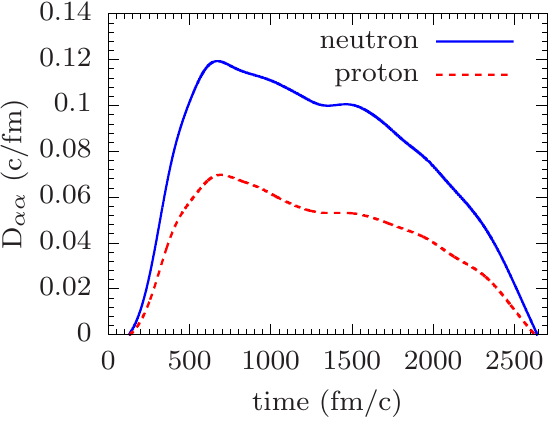}
\caption{Neutron and proton diffusion coefficients as a function of time in ${}^{48} \text{Ca}+{}^{238} \text{U}$ collision at $E_{\text{c.m.}} =193$~MeV for the tip geometry of the uranium and initial orbital angular momentum $\ell=40\hbar$.}
\label{fig1}
\end{figure}

Stochastic part of the drift coefficients, $\delta v_{p}^{\lambda } (t)$ and $\delta v_{n}^{\lambda } (t)$, provide the source for generating fluctuations in mass and charge asymmetry degrees of freedom. According to the SMF approach, stochastic parts of drift coefficients have Gaussian random distributions with zero mean values, $\delta \bar{v}_{p}^{\lambda } (t)=0$, $\delta \bar{v}_{n}^{\lambda } (t)=0$, and the auto-correlation functions of stochastic drift coefficient integrated over their time history determine diffusion coefficients $D_{\alpha \alpha } (t)$ for proton and neutron transfers,
\begin{align}\label{eq4}
\int_{0}^{t}dt'\overline{\delta v_{\alpha }^{\lambda } (t)\delta v_{\alpha }^{\lambda}(t')} =D_{\alpha \alpha } (t)\;.                                                                           \end{align} 
In general diffusion coefficients involve a complete set of particle-hole states. It is possible to eliminate the entire set of particle states by employing closure relations in the diabatic limit, which is a good approximation for evolution of TDHF wave function during short time intervals. This provides a great simplification and as a result, diffusion coefficients are determined entirely in terms of the occupied single-particle states of the TDHF evolution. Explicit expressions of diffusion coefficients are provided in previous publications~\cite{ayik2018,yilmaz2018,ayik2019,sekizawa2020,ayik2020,yilmaz2020} and for analysis of these coefficients please see Appendix B in Ref.~\cite{ayik2018}. As seen in these expressions, the source of fluctuations, which are expressed with diffusion coefficient, are specified by the mean-field properties. This result is consistent with the fluctuation dissipation theorem of non-equilibrium statistical mechanics and greatly simplifies calculations of the diffusion coefficient. Diffusion coefficients include the quantal effects due to shell structure, Pauli blocking, and full effect of the collisions geometry without any adjustable parameters.  We observe that there is a close analogy between the quantal expression and the classical diffusion coefficient in the random walk problem~\cite{gardiner1991,weiss1999,risken1996}. The direct part is given as the sum of the nucleon currents across the window from the target-like fragment to the projectile-like fragment and from the projectile-like fragment to the target-like fragment, which is integrated over the memory. This is analogous to the random walk problem, in which the diffusion coefficient is given by the sum of the rate for the forward and backward steps.  The second part in the quantal diffusion expression stands for the Pauli blocking effects in nucleon transfer mechanism, which does not have a classical counterpart. As an example, in Fig.~\ref{fig1} we present neutron and proton diffusion coefficients in ${}^{48} \text{Ca}+{}^{238} \text{U}$ collision at $E_{\text{c.m.}} =193$~MeV for the initial angular momentum $\ell=40\hbar$.

\subsection{Derivatives of drift coefficients}
\begin{figure}[!t]
\includegraphics*[width=8.6cm]{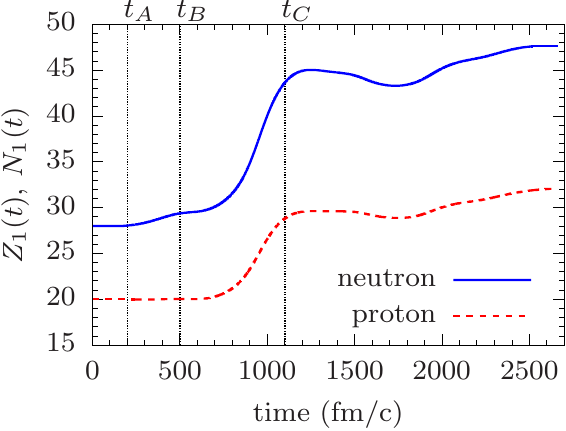}
\caption{Mean drift path of the projectile-like fragments in $(N,Z)-$ plane for the ${}^{48} \text{Ca}+{}^{238} \text{U}$ collision with tip geometry of the uranium at bombarding energy $E_{cm} =193$~MeV and initial angular momentum $\ell=40\hbar$.}
\label{fig2}
\end{figure}

\begin{figure}[!t]
\includegraphics*[width=8.6cm]{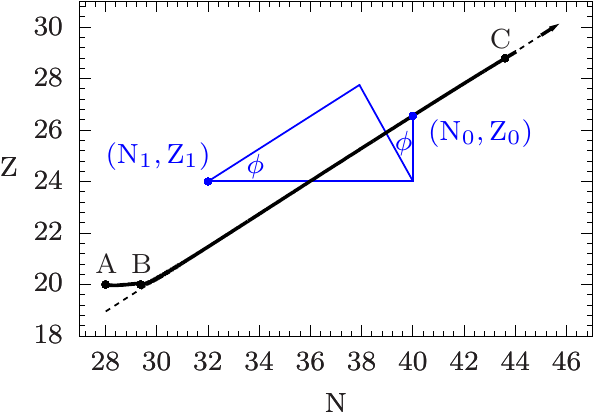}
\caption{The mean-drift path in the $(N,Z)$-plane for the ${}^{48} \text{Ca}+{}^{238} \text{U}$ collision with tip geometry of the uranium at bombarding energy $E_{cm} =193$~MeV and initial angular momentum $\ell=40\hbar$.}
\label{fig3}
\end{figure}

To solve the coupled Langevin Eqs.~\eqref{eq3} we need to evaluate derivatives of the mean drift coefficients with respect to neutron and proton numbers. A single mean-field event is not sufficient to evaluate these derivatives. To calculate derivatives, one needs to evolve several mean-field events with similar initial conditions.  As another possibility for determining the derivatives, we can employ the Einstein relation in the over damped limit~\cite{ayik2018,yilmaz2018,ayik2019,sekizawa2020,ayik2020}. In the over damped limit, drift coefficients are related to the driving potential energy surface in $(N,Z)$-plane as,
\begin{align}\label{eq5}
 v_{n}  &=-\frac{D_{NN} }{T^{*} } \frac{\partial U}{\partial N_{1} }\;,\nonumber\\
 v_{n} &=-\frac{D_{NN} }{T^{*} } \frac{\partial U}{\partial N_{1} }\;,
\end{align}
where $T^{*}$ and $U(N_{1} ,Z_{1} )$ represent the effective temperature and the potential energy surface of the system.  As an example, Fig.~\ref{fig2} shows the evolution of the mean values of neutron and proton numbers of projectile-like fragments as a function of time in ${}^{48} \text{Ca}+{}^{238} \text{U}$ collision at $E_{\text{c.m.}} =193$~MeV for the tip orientation with initial orbital angular momentum $\ell=40\hbar$.  The thick black line in Fig.~\ref{fig3} shows the drift path, which represents the mean evolution of neutron and proton numbers of the projectile-like fragments in the $(N,Z)$-plane. The charge asymmetry of projectile ${}^{48} \text{Ca}$ and target ${}^{238} \text{U}$ are $(N-Z)/(N+Z)\approx 0.17$ and $(N-Z)/(N+Z)\approx 0.23$, respectively. During the initial phase of the collision from touching point at $t_{A} =180$~fm/c until about $t_{B} =480$~fm/c system rapidly evolves toward charge equilibration with charge asymmetry approximately equal to $(N-Z)/(N+Z)\approx 0.20$. After this instant the system follows nearly a straight line path, referred to as the iso-scalar path, toward mass symmetry at which neutron and proton numbers are $N_{0} =(N_{p} +N_{T} )/2=87$, $Z_{0} =(Z_{p} +Z_{T} )/2=56$. We approximately describe the potential energy surface of the the system in $(N,Z)-$ plane in terms of two parabolas. One of these parabolas is oriented along the iso-scalar direction with its  minimum located at the mass symmetry point and the second one is oriented in a direction perpendicular to the iso-scalar path with its minimum located at the iso-scalar path at each point and it is  referred to as the iso-vector path. 
Potential energy of a fragment with $(N_{1} ,Z_{1} )$ is approximately given by,                      
\begin{align}\label{eq6}
U(N_{1} ,Z_{1} )=\frac{1}{2} aR_{S}^{2} (N_{1} ,Z_{1} )+\frac{1}{2} bR_{V}^{2} (N_{1} ,Z_{1} )\;.
\end{align} 
Here, $R_{V} =(Z_{0} -Z_{1} )\sin \phi +(N_{0} -N_{1} )\cos\phi$ and $R_{S} =(Z_{0} -Z_{1})\cos \phi -(N_{0} -N_{1})\sin \phi$, represent the distance of a fragment with $(N_{1} ,Z_{1} )$ from the equilibrium point and the perpendicular distance from the iso-scalar path, respectively. The angle $\phi$ is the angle between the iso-scalar path and $N-$axis, which is about $\phi =32^{\circ}$. Using Einstein relations in Eq.~\eqref{eq5}, we can determine the reduced curvature parameters $\alpha =a/T^{*}$ and $\beta =b/T^{*}$ in terms of drift and diffusion coefficients. Since only ratios of the curvature parameters $(a,b)$ and the effective temperature appear, the effective temperature is not a parameter in the description. Due to the shell effect the reduced curvature parameters vary in time during the TDHF evolution.  We can estimate reduced curvature parameters by taking average over suitable time intervals. Time interval from touching point A at $t_{A} =200$~fm/c until point B at $t_{B} =500$~fm/c, where charge asymmetry equilibration is reached, provides a suitable interval for calculating the average value of the reduced iso-vector curvature parameter, 
\begin{align}\label{eq7}
\alpha =\int _{t_{A} }^{t_{B} }\left(\frac{v_{n} (t)\sin \phi }{D_{NN} (t)} -\frac{v_{p} (t)\cos \phi }{D_{ZZ} (t)} \right) /R_{S} (t)dt\;.
\end{align} 
We can also estimate the reduced iso-scalar curvature parameter as average over the time interval from $t_{B} =500$~fm/c until close to separation time $t_{C} =1200$~fm/c,
\begin{align}\label{eq8}
\beta =\int _{t_{B} }^{t_{C} }\left(\frac{v_{n} (t)\cos \phi }{D_{NN} (t)} +\frac{v_{p} (t)\sin \phi }{D_{ZZ} (t)} \right) /R_{V} (t)dt\;.
\end{align} 
We deduce average value of the reduced curvature parameters as $\alpha =0.13$ and $\beta =0.05$. The potential energy surface has a sharp slope in iso-vector direction and much shallower in the iso-scalar direction. This is a typical behavior of the potential energy surface. In a previous work~\cite{ayik2018}, we have determined the curvature parameters as a function of time for the same system. Here, we evaluate average values of the reduced curvature parameters over the suitable time intervals. In heavy di-nuclear systems centrifugal potential energy has a small contribution to the potential energy surface. Therefore, the curvature parameters we estimated for the initial angular momentum $\ell=40\hbar$ provide a good approximation for the other angular momenta as well. Since the drift coefficients have an analytical form, we can immediately determine their derivatives as,
\begin{align}\label{eq9}
\frac{\partial \nu _{n}}{\partial N_{1} } &=-D_{NN} \left(\alpha \sin ^{2} \phi +\beta \cos ^{2} \phi \right)\\
\frac{\partial \nu _{z} }{\partial Z_{1} } &=-D_{ZZ} \left(\alpha \cos ^{2} \phi +\beta \sin ^{2} \phi \right)\\
\frac{\partial \nu _{n} }{\partial Z_{1} } &=-D_{NN} \left(\beta -\alpha \right)\sin \phi \cos \phi\\
\frac{\partial \nu _{z} }{\partial N_{1} } &=-D_{ZZ} \left(\beta -\alpha \right)\sin \phi \cos \phi\;.
\end{align} 
The curvature parameter $\alpha$ perpendicular to the beta stability valley is much larger than the curvature parameter $\beta$ along the stability valley. Consequently, $\beta$ does not have an appreciable effect on the derivatives of the drift coefficients.

\subsection{Fragment probability distributions}
\begin{figure}[!t]
\includegraphics*[width=8.6cm]{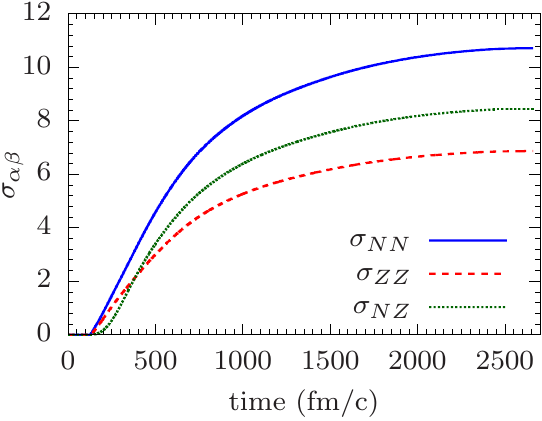}
\caption{Neutron, proton and mixed variances as a function of time in the ${}^{48} \text{Ca}+{}^{238} \text{U}$ collision for the tip geometry of the uranium at bombarding energy $E_{\text{c.m.}} =193$~MeV and initial angular momentum $\ell=40\hbar $.}
\label{fig4}
\end{figure}

In general, joint probability distribution function $P_{\ell}(N,Z)$ for producing a binary fragment with neutron $N$ and proton $Z$ numbers is determined by generating a large number of solutions of Langevin Eqs.~\eqref{eq3}.  It is well known that the Langevin equation is equivalent to the Fokker-Planck equation for the distribution function of the macroscopic variables~\cite{risken1996}. In the particular case when drift coefficients are linear functions of macroscopic variables, as we have in Eq.~\eqref{eq3}, the proton and neutron distribution function for initial angular momentum $\ell$ is given as a correlated Gaussian function described by the mean values, the neutron, proton and, mixed dispersions as,                                  
\begin{align}\label{eq10}
P_{\ell} (N,Z)=\frac{1}{2\pi \sigma _{NN}(\ell)\sigma _{ZZ}(\ell)\sqrt{1-\rho _{\ell}^{2}}}\exp\left(-C_{\ell}\right)\;.
\end{align} 
Here, the exponent $C_{\ell} $ for each impact parameter is given by
\begin{align}\label{eq11}
C_{\ell} =\frac{1}{2\left(1-\rho _{\ell}^{2} \right)} &\left[\left(\frac{Z-Z_{\ell} }{\sigma _{ZZ} (\ell)} \right)^{2} -2\rho \left(\frac{Z-Z_{\ell} }{\sigma _{ZZ} (\ell)} \right)\left(\frac{N-N_{\ell} }{\sigma _{NN} (\ell)} \right)\right.\nonumber\\
&\left.+\left(\frac{Z-Z_{\ell} }{\sigma _{ZZ} (\ell)} \right)^{2} \right]\;,
\end{align} 
with $\rho _{\ell} =\sigma _{NZ}^{2} (\ell)/\left(\sigma _{ZZ} (\ell)\sigma _{NN} (\ell)\right)$. It is possible to deduce coupled differential equations for variances $\sigma _{NN}^{2} (\ell)=\overline{\delta N^{\lambda } \delta N^{\lambda } }$, $\sigma _{ZZ}^{2} (\ell)=\overline{\delta Z^{\lambda } \delta Z^{\lambda } }$, and co-variances $\sigma _{NZ}^{2} (\ell)=\overline{\delta N^{\lambda}\delta Z^{\lambda}}$ by multiplying Langevin Eq.~\eqref{eq3} by $\delta N^{\lambda}$ and $\delta Z^{\lambda }$ and taking the average over the ensemble generated from the solution of the Langevin equation. These coupled equations are presented in Refs.~\cite{ayik2018,yilmaz2018,ayik2019,sekizawa2020,ayik2020}. Variances and co-variances are determined from the solutions of these coupled differential equations with initial conditions $\sigma _{NN}^{2} (t=0)=0$, $\sigma _{NN}^{2} (t=0)=0$, and $\sigma _{NN}^{2} (t=0)=0$ for each angular momentum. As an example, Fig.~\ref{fig4} shows neutron, proton and mixed variances as a function of time for the ${}^{48} \text{Ca}+{}^{238} \text{U}$ collisions in the tip orientation of uranium with the bombarding energy $E_\mathrm{c.m.} =193$~MeV for the initial angular momentum $\ell=40\hbar$. The set of coupled equations are also familiar from the phenomenological nucleon exchange model, and they were derived from the Fokker-Planck equation for the fragment neutron and proton distributions in the deep-inelastic heavy-ion collisions~\cite{schroder1981,merchant1982}. The quantities$N_{\ell} =\bar{N}_{\ell}^{\lambda } $, $Z_{\ell} =\bar{Z}_{\ell}^{\lambda }$ denote the mean neutron and proton numbers of the target-like or project-like fragments. These mean values are determined by the TDHF calculations.

\section{Results}
\label{results}
In a previous investigation~\cite{ayik2018}, we have calculated primary fragment mass yield in ${}^{48} \text{Ca}+{}^{238} \text{U}$ collisions at bombarding energy $E_{\text{c.m.}} =193$~MeV, and compared the result with experimental data of Kozulin~\textit{et al.}~\cite{kozulin2014b}. Here we present results for cross-sections $\sigma(N,Z)$ of production of primary and secondary isotopes with proton numbers $Z=64-80$ in the same system at the same bombarding energy. Also we improve the yield calculation by including binary fragment production by fusion-fission mechanism. We assume the same experimental set up where the detectors are placed at angles $\pm 64^{\circ}$ with $10^{\circ}$ acceptance range in laboratory frame. We consider three collision geometries of the target nucleus: tip configuration where the symmetry axis of uranium is parallel to the beam direction and two side geometries in which symmetry axis of uranium perpendicular to the beam direction (in reaction plane and perpendicular to reaction plane). According to the TDHF calculations, mean trajectories of 
\begin{table}[!h]
    \caption{Final orbital angular momentum $\ell_{f}$, final average total kinetic energy $TKE$, average total excitation energy $E^{*}$ and scattering angles corresponding to a range of initial angular momentum $\ell_{i}$.}
    \label{tab1}
    \begin{ruledtabular}
        \begin{tabular}{|c|c|c|c|c|c|c|}
            $\ell_i$ $(\hbar)$	& $\ell_f$ $(\hbar)$ & $TKE$ & $E^*$ & $\theta_{cm}$ & $\theta_{1}^{lab}$ & $\theta_{2}^{lab}$  \\
            \hline
            38 &32.8	&203.6	&76.1	&92.8	&77.6	&52.7\\
            40 &33.4	&207.8	&78.9	&86.8	&72.0	&57.3\\
            42 &33.2	&203.8	&78.4	&85.3	&70.6	&57.8\\
            44 &34.7	&206.4	&69.4	&91.5	&76.6	&53.6\\
            46 &38.6	&197.7	&71.6	&91.5	&52.4	&75.0\\
            48 &38.0	&195.6	&77.6	&88.9	&74.0	&54.1\\
            50 &39.8	&197.2	&74.6	&86.9	&72.1	&55.6\\
            52 &42.8	&199.5	&72.2	&85.3	&70.7	&56.9\\
            54 &43.6	&196.2	&73.0	&84.1	&69.7	&57.1\\
            56 &44.2	&187.7	&81.5	&83.5	&68.9	&56.6
        \end{tabular}
    \end{ruledtabular}
\end{table}
fragments reach the acceptance range of detectors in collisions only in the tip configuration with initial angular momentum in the range $\ell=38-56\hbar$. Table~\ref{tab1} shows results of TDHF calculations in tip geometry for final angular momentum, final total kinetic energy, total excitation energy, scattering angles in the center of mass frame and laboratory frame for a range initial angular momentum. The results presented in this table are obtained by performing calculations with the TDHF code~\cite{umar2006c}. According to the detector set up, the range of initial angular momentum $\ell=38-56\hbar$ is within the acceptance range of detectors. Table~\ref{tab2} shows the mean values of initial and final mass and charge numbers of projectile-like and target-like fragments. Same table shows asymptotic values of quantum diffusion calculations for neutron $\sigma _{NN}^{2}$, proton $\sigma _{ZZ}^{2}$ and mixed $\sigma _{NZ}^{2}$ variances.  Since we determine curvature parameters of the potential energy using a different approach, some differences appear for dispersion results from those presented in~\cite{ayik2018}.  
         
\subsection{Primary mass yield}
In the previous investigation, we included only the multi-nucleon transfer (mnt) processes in the primary yield calculations. Here, we improve the yield calculations by incorporating binary fragments production in fusion-fission (ff) processes.   We can express the yield of primary fragments mass distribution according to
\begin{align}\label{eq12}
Y(A)=\eta\left(P^{mnt}(A)+P^{ff}(A)\right)\;,
\end{align}
where $P^{mnt}(A)$ and $P^{ff}(A)$ are the fragment mass distributions due to multi-nucleon transfer and due to fusion-fission mechanism, respectively, and $\eta$ is a normalization constant. According to quantal diffusion calculations, probability distribution due to multi-nucleon transfer is given by
\begin{align} \label{eq16} 
P(A)=\frac{1}{\sum _{\ell }\left(2\ell +1\right) } \sum _{\ell}\left(2\ell +1\right)\left[P_{\ell }^{pro}(A) + P_{\ell }^{tar}(A) \right]\;,
\end{align}
\begin{table}[!h]
    \caption{Mean values of mass and charge numbers of the initial and final fragments, neutron variance, proton variance and mixed variance for a range of initial orbital angular momentum $\ell_{i}$.}
    \label{tab2}
    \begin{ruledtabular}
        \begin{tabular}{|c|c|c|c|c|c|c|c|c|}
            $\ell_i$ $(\hbar)$ & $A_1^f$ & $Z_1^f$ & $A_2^f$  & $Z_2^f$ & $\sigma_{NN}^2$ & $\sigma_{ZZ}^2$ & $\sigma_{NZ}^2$ & $\sigma_{AA}^2$\\
            \hline
            38    & 78.5 & 31.5 & 207.5 & 80.5 & 115.7 & 50.5 & 67.8 & 234.0 \\
            40    & 79.7 & 32.1 & 206.3 & 79.9 & 116.8 & 50.9 & 68.6  & 236.3 \\
            42    & 78.6 & 31.6   & 207.4 & 80.4 & 114.6 & 49.9 & 67.1 & 231.6 \\
            44    & 77.5 & 31.2 & 208.5 & 80.8 & 107.2 & 46.8 & 62.3 & 216.3 \\
            46    & 75.0 & 30.1 & 211.0 & 81.9 & 102.4 & 44.8 & 59.2 & 206.4 \\
            48    & 75.8 & 30.5 & 210.2 & 81.5 & 101.6 & 44.4 & 58.6 & 204.6 \\
            50    & 76.1 & 30.6 & 209.9 & 81.4 & 99.8 & 43.7 & 57.5 & 201.0 \\
            52    & 76.3 & 30.8 & 209.7 & 81.3 & 97.4 & 42.7 & 55.9 & 196.0 \\
            54    & 75.2 & 30.3 & 210.8  & 81.7 & 94.5 & 41.4 & 54.0 & 189.9 \\
            56    & 74.2  & 30.1 & 211.8 & 81.9 & 91.6 & 40.2 & 52.1 & 183.9           
        \end{tabular}
    \end{ruledtabular}
\end{table}
with the range of initial angular momentum spanning the interval $\ell=38-56\hbar$.  Here $P_{\ell }^{pro}(A)$ and $P_{\ell }^{tar}(A)$ are mass distributions of projectile-like and target-like fragments. We can determine the probability distribution $P_{\ell }(A)$ of the mass number $A$ of the primary fragments, by integrating Eq.~\eqref{eq16} over $Z$ substituting $N=A-Z$  to find,
\begin{align} \label{eq14} 
P^{\ell}(A)=\frac{1}{\sqrt{2\pi}\sigma_{AA}} \exp \left\{-\frac{1}{2} \left[\left(\frac{A-A_{\ell } }{\sigma _{AA}(\ell) } \right)^{2} \right]\right\}\;.
\end{align}
\begin{figure}[!t]
    \includegraphics*[width=8.6cm,height=7.0cm]{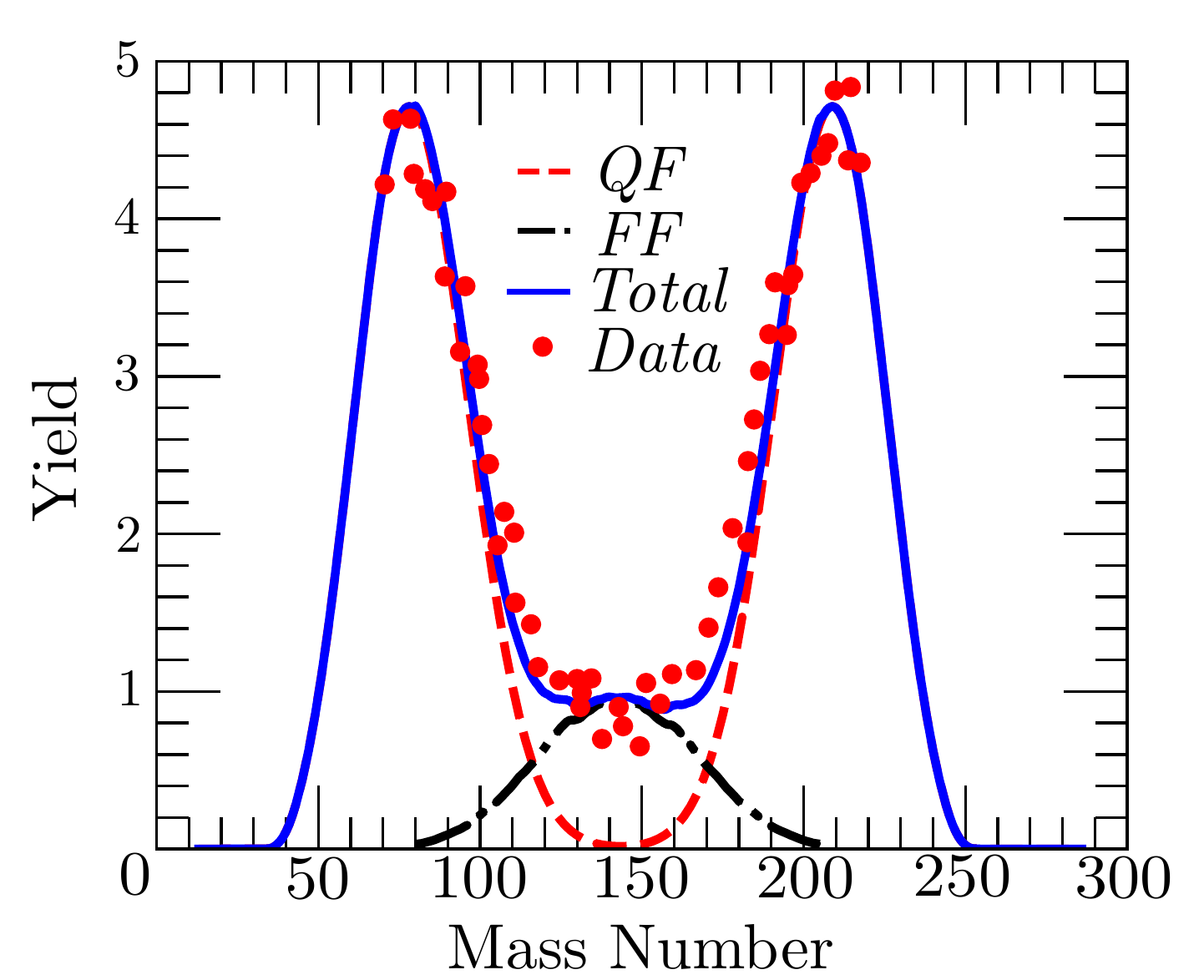}
    \caption{Solid blue line shows the combined primary yield of multi-nucleon transfer (dashed red line in center region) and binary fission (black dashed dotted line) as function fragment mass A in collision of ${}^{48} \text{Ca}+{}^{238} \text{U}$ at $E_{\text{c.m.}} =193.1$~MeV. Red solid points show data taken from~\cite{kozulin2012}.}
    \label{fig5}
\end{figure}
Here, mass dispersion is given by $\sigma _{AA}^{2} =\sigma _{NN}^{2} +\sigma _{ZZ}^{2} +2\sigma _{NZ}^{2}$ and $A_{\ell}$ indicates the mean value of the mass number projectile-like or target-like for each angular momentum. We cannot employ quantal diffusion approach to determine fragment mass distribution due to fusion-fission mechanism. However we can estimate fusion-fission probability in the following manner: Bombarding energy $E_{\text{c.m.}} =193.1$~MeV is very close to fusion barrier located around $E_{bar} =193.8$~MeV. At this bombarding energy, TDHF calculations are not very reliable for determining fusion probability. We assume that at this bombarding energy near central collisions up to critical angular momentum $\ell_c$ lead to fusion. The excitation energy of the compound nucleus is determined as $E^{*}_{C}=E_{\text{c.m.}}-Q_{gg}$, where $Q_{gg}$ is the ground state Q-value of the compound nucleus relative to the initial state. We determine fission probability compound nucleus into binary fragments using GEMINI++ code~\cite{charity2008}. Normalized mass distribution of binary fragments is given by,
\begin{align} \label{eq15} 
P^{ff}(A)=\frac{1}{\sum _{\ell}\left(2\ell +1\right) } \sum _{\ell}\left(2\ell +1\right)P_{\ell }^{ff}(A)\;.
\end{align}
Here, $P_{\ell }^{ff}(A)$  denotes fission probability distribution of the compound nucleus produced in collision with initial angular momentum $\ell$.  Summation interval extends up to a critical angular momentum $\ell=0-\ell_c$. We expect that average value of fission fragment distribution not to be very sensitive to the magnitude of the critical angular momentum. Hence, in calculations we take the critical angular momentum for leading to fusion as $\ell_c=5\hbar$.  In Fig.~\ref{fig5} solid blue line shows the combined primary yield of multi-nucleon transfer (dashed blue line in center region) and binary fission (blue dashed dotted line) as function of fragment mass. Calculations are compared with data of reference~\cite{kozulin2012} shown by red solid points. Normalization constant in Eq.~\eqref{eq12} is determined by fitting data at a suitable point to give $\eta=214$.
\begin{figure*}[!htb]
    \includegraphics*[width=16cm]{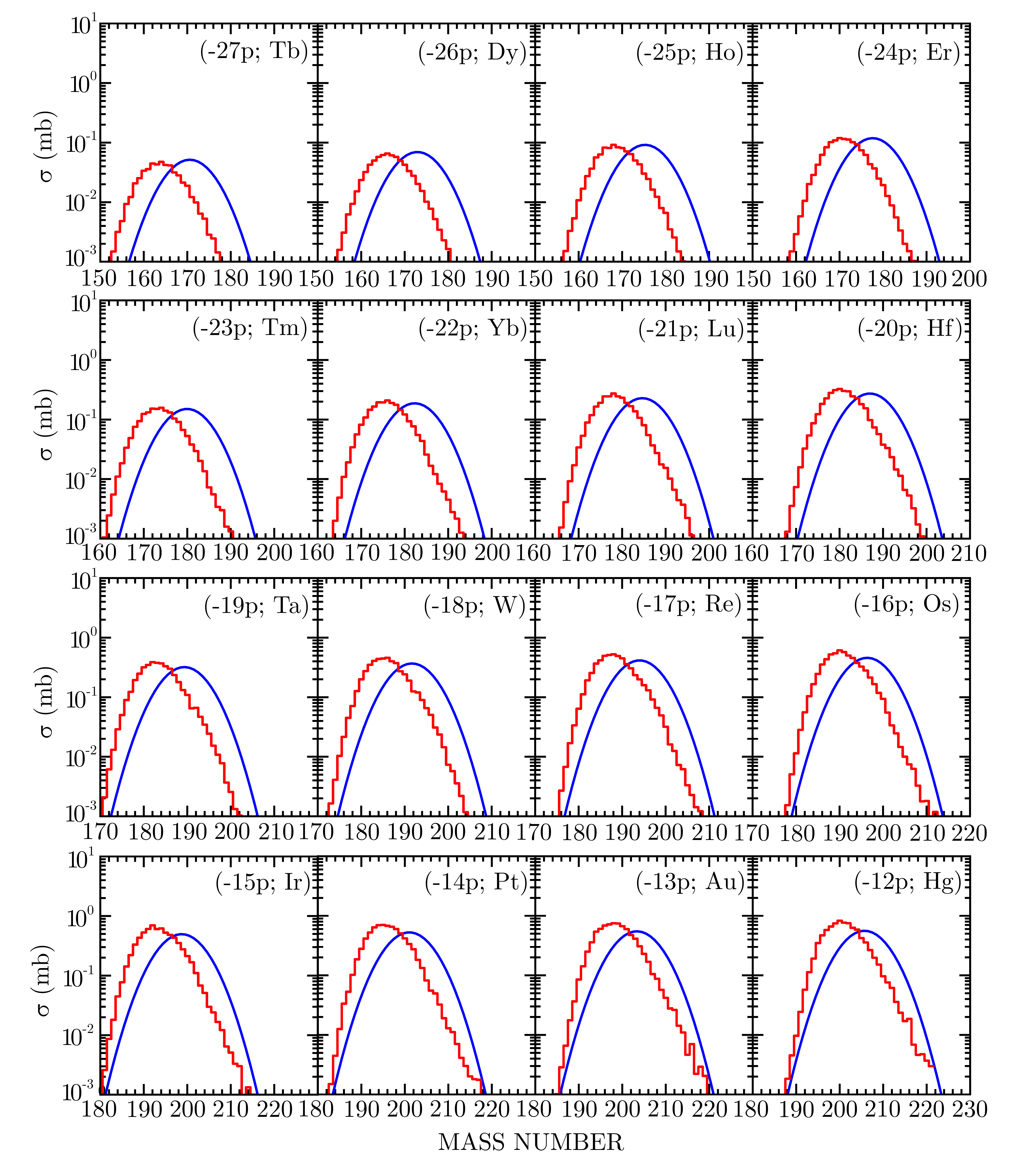}
    \caption{Solid blue lines show primary isotope yields for $Z=65-80$ as function fragment mass $A$ in collision of ${}^{48} \text{Ca}+{}^{238} \text{U}$ at $E_{\text{c.m.}} =193.1$~MeV. Red histograms show production cross-sections of secondary isotopes calculated with GEMINI++.}
    \label{fig6}
\end{figure*}
\subsection{Isotope production with $Z=64-80$}
We calculate the cross sections for production of primary isotopes using the standard expression,
\begin{align}
\sigma (N,Z)=\frac{\pi \hbar ^{2} }{2\mu E_{\text{c.m.}} } \sum _{\ell}(2\ell+1)\left[P_{\ell}^{pro} (N,Z)+P_{\ell}^{tar} (N,Z)\right]\;.
\end{align}   
Here, $P_{\ell}^{pro} (N,Z)$ and $P_{\ell}^{tar} (N,Z)$ denote the probability of producing projectile-like and target-like fragments. These probabilities are given by Eq.~\eqref{eq10} with mean values of projectile-like and target-like fragments, respectively.  In the summation over $\ell$, the range of initial angular momentum depends on the detector geometry in the laboratory frame. In calculations we take the range of angular momentum $\ell=38-56\hbar$ according to the experimental set up of Kozulin~\textit{et al.}~\cite{kozulin2014b}, for the ${}^{48} \text{Ca}+{}^{238} \text{U}$ system at bombarding energy $E_{\text{c.m.}} =193.1$~MeV. 
Blue solid lines in Fig.~\ref{fig6} show the production cross-sections of primary target-like isotopes with proton number $Z=65-80$ as a function of mass number $A$. These primary isotopes are excited and cool down by emitting light particles, mostly neutrons, protons, and alpha particles, and may also decay via binary fission. We analyze de-excitation mechanism of primary fragments using statistical code GEMINI++~\cite{charity2008}. We estimate the total excitation energy of binary primary fragments according to $E_{\ell}^{*}(Z,A)=E_{\text{c.m.}}-\text{TKE}_{\ell}-Q_{gg}(Z,A)-\Delta V_{\ell}(Z)$.  In this expression $\text{TKE}_{\ell}$ is the mean value of total asymptotic kinetic energy in collision with initial angular momentum $\ell$ and $Q_{gg}(Z,A)$ denotes ground state  $Q$-value of the primary binary fragments relative to the initial value. It is possible to add Coulomb correction to the total final kinetic energy due to proton transfer in excess of the mean number of proton transfer~\cite{simenel2010}. We ignore this correction in the present study. We share the total excitation energy and total angular momentum transfer in proportion to the mass ratio of binary primary fragments. Red histograms in Fig.~\ref{fig6} show the secondary cross sections for target-like isotopes with proton numbers $Z=65-80$ as a function of mass number $A$. We note that secondary isotope distributions shift toward the valley of stability by emitting $5-6$ neutrons depending on excitation energy of primary fragments. The secondary cross-sections are not obtained from the primary cross-sections by merely shifting the primary cross-sections with the number of emitted neutrons. According to the statistical de-excitation code GEMINI++, in addition to neutron emission, the secondary isotopes with small proton numbers are populated by de-excitation of primary fragments with larger proton numbers by emitting alpha particles and protons, and by induced secondary fission. The de-excitation mechanisms other than neutron emission increase the secondary cross-sections to the left of peak values of the primary cross-sections, and even overshoot the maximum value of the primary cross-sections values as seen in Fig.~\ref{fig6}. We also note production cross-sections for heavy neutron rich isotopes of mercury, gold and platinum with neutron numbers around $125-130$ are in the order of several micro barns. These cross-sections are much larger than the estimates given by Adamian~\textit{et al.}~\cite{adamian2010c}. 

\section{Conclusions}
\label{conc}
As an extension of previous work, we have investigated multi-nucleon transfer mechanism in ${}^{48} \text{Ca}+{}^{238} \text{U}$ collisions at $E_{\text{c.m.}} =193.1$~MeV using quantal diffusion description based on the SMF approach.  In this approach, transport coefficients associated with macroscopic variables such as charge and mass asymmetry variables are evaluated in terms of time-dependent single-particle wave functions of TDHF theory. Transport description includes quantal effects due to shell structure, full geometry of the collision dynamics, and the Pauli exclusion principle without any adjustable parameters aside from the standard description of the effective Hamiltonian of TDHF theory. Joint probability distribution of primary fragments is determined by a correlated Gaussian function in terms of mean values of neutron-proton numbers and neutron, proton, mixed dispersions for each initial angular momentum.  We calculate the yield of primary fragments as a function of mass number and compare with data of Kozulin~\textit{et al.}. Primary yield contains fragments produced by multi-nucleon transfer and also secondary fission products. Since the bombarding energy is very close to fusion barrier located at about $E_{\text{bar}} =193.8$~MeV, TDHF is not very reliable for determining the fusion mechanism. We assume ${}^{48} \text{Ca}+{}^{238} \text{U}$ collisions lead to fission at near central collisions and determine fission probability employing GEMINI++ code. Calculations produce good description of experimental mass yield distribution. We investigate de-excitation mechanism of heavy target-like primary fragments in the range of $Z=65-80$, and calculate secondary isotope cross-sections as a function of mass number. Calculations predict cross-sections on the order of several micro barns for heavy neutron rich isotopes of mercury, gold and platinum with neutron numbers in the range of $125-130$.   

\begin{acknowledgments}
Authors are thankful to K. Sekizawa for useful discussion on GEMINI++ code. S.A. gratefully acknowledges Middle East Technical University for warm hospitality extended to him during his visits. S.A. also gratefully acknowledges F. Ayik for continuous support and encouragement. This work is supported in part by US DOE Grants No. DE-SC0015513, and in part by US DOE Grant No. DE-SC0013847.
\end{acknowledgments}

\bibliography{VU_bibtex_master}
\end{document}